\documentclass[twoside,11pt,a4paper,final]{article}

\usepackage[dvips]{graphicx}
\usepackage{epsfig}

\usepackage{graphicx}
\usepackage{psfrag}
\usepackage{amsfonts}
\usepackage{colordvi}

\begin{document}

\bibliographystyle{plain}

\begin{center}
\Large{\textbf{Noisy qutrit channels}}\\

\normalsize{Agata Ch\c{e}ci\'{n}ska$^1$, Krzysztof W\'{o}dkiewicz$^{1,2}$}\\

\vspace{1cm}
\footnotesize{
$^1$ Instytut Fizyki Teoretycznej, Uniwersytet Warszawski, Ho\.za 69, 00-681 Warszawa, Poland\\
$^2$ Department of Physics and Astronomy, University of New Mexico, 800 Yale Blvd. NE, Albuquerque, NM 87131, USA
}
\end{center}

\vspace{1cm}
\begin{center}
\textbf{Abstract}
\end{center}

\small{
We present an analysis of spontaneous emission in a 3-level atom as an example of a qutrit state under the action of noisy quantum channels. We choose a 3-level atom with V-configuration to be the qutrit state. Gell-Mann matrices and a generalized Bloch vector (8-dimensional) are used to describe the qutrit density operator. Using the time-evolution equations of atomic variables we find the Kraus representation of spontaneous emission quantum channel (SE channel). Furthermore, we consider a generalized Werner state of two qutrits and investigate the separability condition. We give similar analysis of spontaneous emission for qubit channels. The influence of spontaneous emission on the separability of Werner states for qutrit and qubit states is compared.} \\

\vspace{1cm}
\section{Introduction}

Quantum engineering, teleportation and the idea of constructing quantum computers were brought into focus of recent, widespread, scientific research. Consequently, investigation of quantum phenomena, such as entanglement, has been intensified \cite{1,2}. One of the possible ways is to investigate atomic systems, their entanglement and behaviour under action of quantum channels. Since these might be useful, it is desirable to quantify the influence of channels on states separability and entanglement. The state and channel are very general concepts, hence we would like to focus on particular examples of a spontaneous emission channel involving a quantum state described by a 3-level atom. Spontaneous emission is a proccess that might destroy mutual entanglement. The majority of papers have been devoted to qubit states. It is the purpose of this paper to concentrate on a qutrit state, which has a physical realization in form of a three-level atom in the presence of spontaneous emission. Such a system ia an example of qutrit channels with noise. We compare the qutrit case with the qubit, in other words, three-level and two-level atoms under the action of spontaneous emission channel. \\
The paper is structured as follows: we present description of (single) qutrit states (analogy to Bloch formalism), we choose 3-level atoms with V-configuration to be the qutrit state; equations for atomic evolution in presence of spontaneous emission are given, channel mathematics is presented, with focus on Kraus representation; Kraus operators for spontaneous emission channel (SE channel) are evaluated; analysis of two qutrit state is brought into discussion and generalized Werner state is described together with separability condition; the action of spontaneous emission channel on two qutrit state is investigated; we give Kraus representation for analogous qubit channel and analyze Werner state of two qubits; we compare both cases: qutrit and qubit separability behaviuor under influence of spontaneous emission and recognize that, depending on channel characteristic, the qutrit systems can preserve entanglement longer.

\section{Qutrit state}

The concept of a qubit \cite{3,4} i.e, a quantum state
living in two dimensional Hilbert space is used as a basic
building block of Quantum Information.  Within the framework of
atomic physics   two-level atom is the simplest aphysical
realization of a qubit \cite{3}. Qubit and qubit channels have been investigated comprehensively \cite{3,4,12} and generalization to N-dimensional cases involving qudits has been studied \cite{13,14,15}, though less exhaustively. From physical point of view the use of more complex atomic structures might be advantageous \cite{10}, therefore three-level atoms, qutrits, might deserve more interest and studies. To give their mathematical description, we note that the Bloch formalism of a two-level atom is based on the $SU(2)$ generators given by Pauli matrices, as a basis for qubit density operator \cite{5,6}:
\begin{equation}
\rho_{qubit}=\frac{1}{2}(\mathbb{I}+\vec{b}\cdot\vec{\sigma}),
\end{equation}
   
where $\vec{b}$ is a three dimensional (real) Bloch vector \cite{13}. The mathematical description of a qutrit density operator involves in a natural way the $SU(3)$ generators , called the Gell-Mann matrices $\lambda_i$ \cite{6,16}:
\begin{equation}
\rho=\frac{1}{3}(\mathbb{I}+\sqrt{3}\vec{n}\cdot\vec{\lambda}),
\end{equation} 
where $\vec{n}$ is a real eight dimensional generalized Bloch vector.\\
Qutrit states belong to three-dimensional complex Hilbert space $\mathcal{H}^{(3)}$; pure qutrit states correspond to vectors that satisfy \cite{16}:
\begin{equation}
\vec{n}\cdot\vec{n}=1,\  \vec{n}*\vec{n}=\vec{n},
\end{equation} 
where $(\vec{A}*\vec{B})_k \equiv d_{klm}A_l B_m$, with $d_{klm}$ being a totally symmetric tensor associated with the $SU(3)$ \cite{6}.\\
These two conditions define a generalized Bloch sphere for qutrits, in analogy to Bloch qubit sphere. In general, pure state can be parameterized as follows (in the atomic basis $|1\rangle, |2\rangle, |3\rangle$):
\begin{equation}
|\Psi\rangle = \sin\frac{\xi}{2}\cos\frac{\theta}{2}|1\rangle+e^{\imath\phi_{12}}\sin\frac{\xi}{2}\sin\frac{\theta}{2}|2\rangle+e^{\imath\phi_{13}}\cos\frac{\xi}{2}|3\rangle.
\end{equation}
 Orthogonal states in $\mathcal{H}^{(3)}$ do not correspond to opposite points on $\mathcal{S}^7$ (the seven-dimensional unit sphere in $\mathcal{R}^8$), but to points of maximum opening angle of $\frac{2\pi}{3}$. Distribution of points on $\mathcal{S}^7$ that represent physical states, the generalized Bloch sphere, is highly nontrivial \cite{5,6,17}.

\section{3-level atoms}

We will consider a particular physical realization of a qutrit state, namely 3-level atoms. There are three configurations of 3-level atoms \cite{18} - we choose the so called V-configuration in which the only allowed transitions $|2\rangle \longrightarrow |1\rangle$ and $|3\rangle \longrightarrow |1\rangle$ are depicted on \textit{Fig.1}:

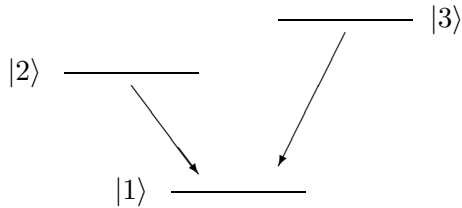
\begin{figure}[h!]
\begin{center}
\begin{picture}(200,110)
\put(140,75){\vector(-1,-2){25}}
\put(60,55){\vector(3,-4){25}}
\put(35,60){\line(1,0){50}}
\put(115,80){\line(1,0){50}}
\put(75,15){\line(1,0){50}}

\put(20,60){\makebox(0,0){$|2\rangle$}}
\put(178,80){\makebox(0,0){$|3\rangle$}}
\put(60,15){\makebox(0,0){$|1\rangle$}}

\end{picture}
\caption{Transitions allowed in 3-level atom with V-configuration.}
\end{center}
\label{fig1}
\end{figure}

The Bloch vector $\bar{n}$ can be expressed by  atomic  populations and dipole transitions:

\[ \begin{array}{ccccccc}
n_1 & = & \frac{\sqrt{3}}{2}(d_1^*+d_1), & \  & n_2 & = & \frac{\sqrt{3}\imath}{2}(d_1^*-d_1),\\
n_4 & = & \frac{\sqrt{3}}{2}(d_2^*+d_2), & \  & n_5 & = & \frac{\sqrt{3}\imath}{2}(d_2^*-d_2),\\
n_6 & = & \frac{\sqrt{3}}{2}(d_3^*+d_3), & \  & n_7 & = & \frac{\sqrt{3}\imath}{2}(d_3^*-d_3),\\
n_3 & = & \frac{\sqrt{3}}{2}(1-2p_2-p_3), & \  & n_8 & = & \frac{1}{2}(1-3p_3).
\end{array}\]

(recall that $ d_1: |1\rangle \rightarrow |2\rangle,\   d_2: |1\rangle \rightarrow |3\rangle,\   d_3: |2\rangle \rightarrow |3\rangle.$)
 
\subsection*{Evolution in the presence of spontaneous emission}

Spontaneous emission is a dissipative process, in which the atom is coupled to the electromagnetic vacuum.
The dissipative  evolution of the atomic  variables in the presence of
spontaneous emission is  characterized by two Einstein coefficients $A_1$ and $A_2$ that describe the
irreversible depopulation from excited states. This corresponds to the following Bloch equations
with time independent coefficients:
\begin{equation}
\frac{d}{dt}\vec{n}(t)=\mathcal{M}\vec{n}(0)+\vec{m}_0,
\end{equation}
where $\mathcal{M}$ is a (almost) diagonal matrix with only one non-diagonal entry and $\vec{m}_0$ is a translation.\\
The solution to this equation is of the form:

\begin{equation}
\vec{n}(t) = \mathcal{T}\cdot\vec{n}(0) + \vec{n}_0(t),
\end{equation}

\small{
\begin{equation}
\mathcal{T} = \left(\begin{array}{cccccccc}

e^{-\frac{A_1t}{2}}&0&0&0&0&0&0&0\\
0&e^{-\frac{A_1t}{2}}&0&0&0&0&0&0\\
0&0&e^{-A_1t}&0&0&0&0&\frac{1}{\sqrt{3}}(e^{-A_2t}-e^{-A_1t})\\
0&0&0&e^{-\frac{A_2t}{2}}&0&0&0&0\\
0&0&0&0&e^{-\frac{A_2t}{2}}&0&0&0\\
0&0&0&0&0&e^{-\frac{(A_1+A_2)t}{2}}&0&0\\
0&0&0&0&0&0&e^{-\frac{(A_1+A_2)t}{2}}&0\\
0&0&0&0&0&0&0&e^{-A_2t}
\end{array}\right)\end{equation}

where $\vec{n}_0(t)$ is a translation dependant on time:
\begin{equation}
\vec{n}_0(t) = \left(\begin{array}{c}
0\\
0\\
\frac{1}{2\sqrt{3}}(3-e^{-A_2t}-2e^{-A_1t})\\
0\\
0\\
0\\
0\\
\frac{1}{2}(1-e^{-A_2t})
\end{array}\right)
\end{equation}}
\normalsize
Hence, density operator representing the state of an atom in presence of spontaneous emission is of the form:
\begin{equation}
\rho(t)=\frac{1}{3}\left(\mathbb{I}+\sqrt{3}(\mathcal{T}\vec{n}(0)+\vec{n}_0(t))\cdot\vec{\lambda}\right).
\end{equation}

\section{Completely positive maps and Kraus representation}

Channel acting on a density operator maps density operators into density operators\cite{3,12,19}:

\begin{equation}
\Phi: \rho_{in}\ \mapsto \rho_{out} 
\end{equation}

It is well known that  the channel transformation $\Phi$   is described by  a completely positive map
(CPM) \cite{12}. The simplest way to describe a channel is by means of an operator-sum representation
\cite{19,20}:
\begin{equation}
\rho_{out}=\Phi(\rho_{in})=\sum_i \mathcal{K}_i\rho_{in}\mathcal{K}_i^{\dagger},
\end{equation}
where $\mathcal{K}_i$ are Kraus operators that satisfy normalization condition:
\begin{equation}
\sum_i \mathcal{K}_i^{\dagger}\mathcal{K}_i=\mathbb{I}.
\end{equation}
From the Bloch equations, we can calculate the action of spontaneous emission channel (\textit{SE}
channel) on the V-atom in terms of the  operator-sum representation. In this case the set of corresponding
Kraus operators is as follows:

 \begin{equation}
\begin{array}{ccccc}
\mathcal{K}_0 =\left(\begin{array}{ccc}1&0&0\\0&e^{-\frac{A_1t}{2}}&0\\0&0&e^{-\frac{A_2t}{2}}\end{array}\right),&\  &
\mathcal{K}_1 =\left(\begin{array}{ccc}0&\sqrt{1-e^{-A_1t}}&0\\0&0&0\\0&0&0\end{array}\right),&\  &
\mathcal{K}_2 =\left(\begin{array}{ccc}0&0&\sqrt{1-e^{-A_2t}}\\0&0&0\\0&0&0\end{array}\right).
\end{array}\end{equation}

\section{Influence of SE channel on state separability}

\subsection{Generalized Werner state for two qutrits}
We generalize our discussion to the situation in which we have two qutrits. The generalized Werner state describing two qutrits labelled by A and B is of the form \cite{5}:
\begin{equation}
\rho_{\epsilon}=\frac{1-\epsilon}{9}\mathbb{I}^A\otimes\mathbb{I}^B+\epsilon|\Psi^{AB}\rangle\langle\Psi^{AB}|,
\end{equation}
where $0\leq \epsilon \leq 1$. The state is a convex combination of a maximally mixed state and a pure state. The aim is to characterize the values of parameter $\epsilon$ for which $\rho_{\epsilon}$ is separable \cite{8,9} (meaning it can be represented as an ensamble of product states).\\
We will consider only Werner states that consist of a specific pure state, namely:
\begin{equation}
|\Psi\rangle = \frac{1}{\sqrt{3}}(|1^A\rangle\otimes|1^B\rangle+|2^A\rangle\otimes|2^B\rangle+|3^A\rangle\otimes|3^B\rangle).
\end{equation}
To investigate the separability condition on $\epsilon$ we follow the discussion given in \cite{5}. Therefore we represent density operator $\rho_{\epsilon}$ in the basis of  the Gell-Mann matrices $\{\lambda_{\alpha}\}_{\alpha=0}^8$ enriched by $\lambda_0=\sqrt{\frac{2}{3}}\mathbb{I}$ (with $tr(\lambda_{\alpha}\lambda_{\beta})=2\delta_{\alpha\beta}$):

\begin{equation}
\rho_{\epsilon}=\frac{1}{9}c_{\alpha\beta}\lambda^A_{\alpha}\otimes\lambda^B_{\beta},\   c_{\alpha\beta}=\frac{9}{4}tr\{\rho\lambda^A_{\alpha}\otimes\lambda^B_{\beta}\}.
\end{equation}  

In this form, the state is characterized by the $c_{\alpha\beta}$ coefficients ($\alpha,\beta\in\{0,...,8\}$). It turns out \cite{5} that the condition for the state to be separable is of the form: 

\begin{equation}
4\epsilon=\frac{1}{3}\sum_{j=1}^8|c_{jj}| \le 1.\end{equation}
Hence we obtain:
\begin{equation}
\epsilon\le\frac{1}{4}.\end{equation}

\subsection{SE channel action on qutrits}

\begin{figure}[h!]
\psfrag{s}{s(t)}
\psfrag{t}{t}
\psfrag{a}{\footnotesize{1/4}}
\psfrag{b}{\footnotesize{$s_{qt}(A_1=2,A_2=4)$}}
\psfrag{c}{\footnotesize{$s_{qt}(A_1=A_2=4)$}}

\begin{center}

\includegraphics[scale=1.2]{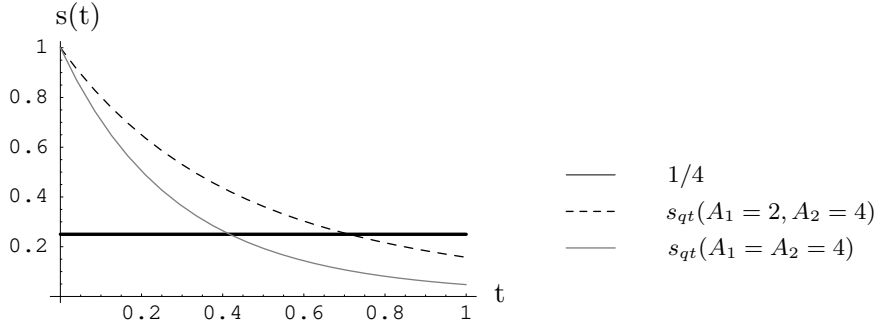}
\caption{The function $s_{qt}(t)=\frac{\epsilon}{8}\left(2e^{-\frac{1}{2}A_1t}+2e^{-\frac{1}{2}A_2t}+2e^{-\frac{1}{2}(A_1+A_2)t}+e^{-A_1t}+e^{-A_2t}\right)$ for $\epsilon=1$. Region below $s=\frac{1}{4}$ corresponds to separable states.}
\label{fig2}
\end{center}
\end{figure}

Action of the channel $\Phi$ on $\rho_{\epsilon}$ changes the coefficients:
\begin{equation}
\Phi: c_{\alpha\beta}\  \mapsto c_{\alpha\beta}(t),
\end{equation}
Therefore, the condition on the $\epsilon$ to produce a separable state becomes time dependant.\\
Consider the channel that alters only one subsystem (for instance $A$):
\begin{equation}
\Phi_1(\rho_{\epsilon})=\sum_{i=0}^2 (\mathcal{K}^A_i\otimes\mathbb{I}^B)\rho_{\epsilon}(\mathcal{K}^A_i\otimes\mathbb{I}^B)^{\dagger}=\frac{1}{9}c_{\alpha\beta}(t)\lambda_{\alpha}\otimes\lambda_{\beta}.
\end{equation}
In this case the separability condition is:
\begin{equation} s_{qt}(t)\equiv\frac{\epsilon}{8}\left(2e^{-\frac{1}{2}A_1t}+2e^{-\frac{1}{2}A_2t}+2e^{-\frac{1}{2}(A_1+A_2)t}+e^{-A_1t}+e^{-A_2t}\right)\leq\frac{1}{4},
\end{equation}
where obviously, for $t=0$ we have:
\begin{equation}
s_{qt}(0)=\epsilon\leq \frac{1}{4}.
\end{equation}
The function $s_{\mathrm{qt}}(t)$ is shown on \textit{Fig.\ref{fig2}}. It is clear, that even initially maximally entangled state (meaning $\epsilon=1$) becomes separable eventually. Two qutrit SE channels with various parameters $A_1,A_2$ are compared. 

In the following we shall compare the decoherence of a qutrit with a similar decoherence for a qubit.

\subsection{SE channel and Werner state for qubits}

The Kraus representation for spontaneous emission channel for qubits is as follows ($A$ is the Einstein coefficient):

\footnotesize{
\begin{equation}
\mathcal{K}_0=\left(\begin{array}{cc} 1&0\\ 0& e^{-\frac{At}{2}}\end{array}\right),\  \mathcal{K}_1=\sqrt{1-e^{-At}}\left(\begin{array}{cc}0&1\\0&0\end{array}\right).\end{equation}
}
\normalsize
Whereas the Werner state describing two non-interacting qubits is of the form \cite{21} ($\alpha,\beta\in\{0,1,2,3\}$, $\sigma_0=\mathbb{I}$):

\begin{equation}
\rho_{\epsilon}=\frac{1-\epsilon}{4}\mathbb{I}^A\otimes\mathbb{I}^B+\epsilon|\Psi\rangle\langle\Psi|=\frac{1}{4}d_{\alpha\beta}\sigma_{\alpha}\otimes\sigma_{\beta},
\end{equation}
where:
\begin{equation}
|\Psi\rangle=\frac{1}{\sqrt{2}}(|1\rangle^A\otimes|2\rangle^B+|2\rangle^A\otimes|1\rangle^B).
\end{equation}
The state $\rho_{\epsilon}$ is separable when \cite{5,8}:
\begin{equation}
\epsilon \leq \frac{1}{3}.
\end{equation}

\subsection{SE channel altering qubit separability}

\begin{figure}[h!]
\psfrag{s}{s(t)}
\psfrag{t}{t}
\psfrag{a}{\footnotesize1/3}
\psfrag{b}{\footnotesize$s_{qb}(A=2)$}
\psfrag{c}{\footnotesize$s_{qb}(A=3)$}
\psfrag{d}{\footnotesize$s_{qb}(A=4)$}
\begin{center}
\includegraphics[scale=1.2]{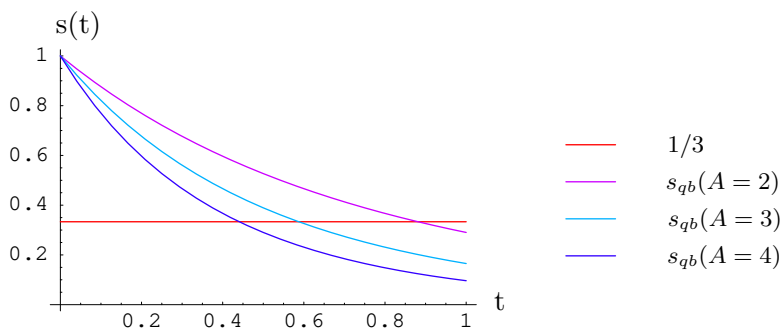}
\caption{Function $s_{qb}(t)=\frac{\epsilon}{3}(2e^{-\frac{At}{2}}+e^{-At})$ for  $\epsilon=1$. Region below $s=\frac{1}{3}$ corresponds to separable states.}
\label{fig3}
\end{center}
\end{figure}

In analogy to what has been done before, we consider SE channel altering only one qubit, with (two) Kraus operators of the form: $\mathcal{K}^A_i\otimes\mathbb{I}^B$. Initial qubit state has the form: 
\begin{equation}
\rho_{\epsilon}=\frac{1}{4}d_{\alpha\beta}\sigma_{\alpha}\otimes\sigma_{\beta}.
\end{equation}
Action of the channel changes coefficients of expansion:
\begin{equation}
\Phi(\rho_{\epsilon})=\rho_{\epsilon}(t)=\sum_{i=1,2}\mathcal{K}_i^{A}\otimes\mathbb{I}^B\rho_{\epsilon}(\mathcal{K}_i^{A}\otimes\mathbb{I}^B)^{\dagger}=\frac{1}{4}d_{\alpha\beta}(t)\sigma_{\alpha}\otimes\sigma_{\beta}.
\end{equation}
And the separability condition leads to the inequality:

\begin{equation}
s_{\mathrm{qb}}(t)\equiv \frac{\epsilon}{3}(2e^{-\frac{At}{2}}+e^{-At})\leq \frac{1}{3},\end{equation}
with initial condition: $ s_{\mathrm{qb}}(0)=\epsilon$.\\ 
Function $s_{qb}(t)$ is depicted on \textit{Fig.\ref{fig3}}.
Again, it is clear that any state becomes eventually separable. Time in which maximally entangled state becomes separable can be calculated and depends on the value of parameter A. 

\subsection{Comparison of qubit and qutrit states under the action of SE channels}

Knowing how spontaneous channel acts on both qutrit and qubit states we can compare these two cases in order to state whether qutrit or qubit Werner states preserve entanglement longer. In \textit{Fig.4} we show the comparison of the SE channel action. Points of intersection of $s_{qt}=\frac{1}{4}$ and $s_{qb}=\frac{1}{3}$, associated with time needed to reach separability, are crucial to this analysis. The proper choice of parameters $A,A_1,A_2$ can lead us to the case in which qutrit entanglement is stronger or vice versa. The function $s_{qt}(t)$ is symmetric with respect to the change $A_1 \leftrightarrow A_2$, the relative value of A parameter (with respect to $A_1,A_2$) in $s_{qb}(t)$ influences the result of qubit-qutrit comparison.   

\begin{figure}[h!]
\psfrag{s}{s(t)}
\psfrag{t}{t}
\psfrag{a}{\footnotesize{1/3}}
\psfrag{b}{\footnotesize{1/4}}
\psfrag{c}{\footnotesize{$s_{qb}(A=2)$}}
\psfrag{d}{\footnotesize{$s_{qb}(A=3)$}}
\psfrag{e}{\footnotesize{$s_{qb}(A=4)$}}
\psfrag{f}{\footnotesize{$s_{qt}(A_1=2,A_2=4)$}}
\begin{center}

\includegraphics[scale=1.2]{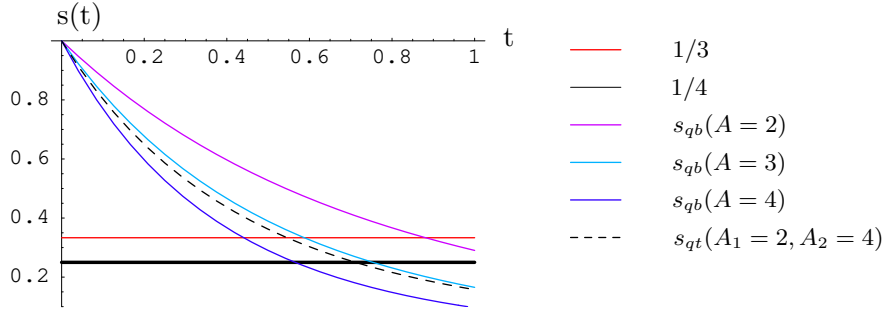}
\caption{Comparison of functions $s_{qt}(t)=\frac{\epsilon}{8}\left(2e^{-\frac{1}{2}A_1t}+2e^{-\frac{1}{2}A_2t}+2e^{-\frac{1}{2}(A_1+A_2)t}+e^{-A_1t}+e^{-A_2t}\right)$ and $s_{qb}(t)=\frac{\epsilon}{3}(2e^{-\frac{At}{2}}+e^{-At})$ for initial condition $\epsilon=1$.}
\label{fig4}
\end{center}
\end{figure}

\section{Summary}

We have presented an example of a qutrit state, namely a 3-level atom in the V configuration, and its evolution under the action of spontaneous emission channel. Separability of two qutrit states is, obviously, influenced by spontaneous emission. This influence is governed by channel parameters. As a consequence, Werner qutrit states can preserve entanglement longer than Werner qubit states - depending on relative values of qubit and qutrit channel parameters. This result might be of some experimental importance when it comes to use of N-level atoms and multipartite entanglement. We plan to investigate further examples of qutrit channels and their influence on state separability. We aim as well at general description of qutrit channels with respect to complete positivity.

\subsection*{Acknowledgements}

This paper was supported by a Polish MNiSW grant No. 1P03B13730.

\end{document}